\documentstyle[12pt]{article}
\def\av#1{\langle #1 \rangle}
\def\half{{\textstyle{1\over 2}}}
\def\cZ{{\cal{Z}}}
\def\cW{{\cal{W}}}
\def\cL{{\cal{L}}}

\def\t12{\theta_{12}}

\def\lapp{{\ \lower 0.6ex \hbox{$\buildrel<\over\sim$}\ }}
\def\gapp{{\ \lower 0.6ex \hbox{$\buildrel>\over\sim$}\ }}
\def\beq{\begin{equation}}
\def\eeq{\end{equation}}
\def\beqn{\begin{eqnarray}}
\def\eeqn{\end{eqnarray}}
\def\to{\rightarrow}

\def\qq{q\bar{q}}

\def\pp{{\rm p\bar{p}}}

\def\tt{t\bar{t}}

\def\GeV{{\rm GeV}}

\def\TeV{{\rm TeV}}

\def\ee{e^+e^-}

\begin{document}
\begin{titlepage}
\vspace*{-1cm}
\begin{flushright}
DTP/94/14   \\
LU-TP--94--3 \\
March 1994 \\
\end{flushright}
\vskip 1.cm
\begin{center}
{\Large\bf
Gluon Radiation and Energy Losses\\[2mm]
 in Top Quark Production}
\vskip 1.cm
{\large Yu.L. Dokshitzer}\footnote{permanent address:
Petersburg Nuclear Physics Institute,
     Gatchina, St.Petersburg 188350, Russia}
\vskip .2cm
{\it Department of Theoretical Physics, University of Lund \\
S\"olvegatan 14A, S-22362 Lund, Sweden }\\
\vskip .4cm
{\large V.A. Khoze}$^1$
\vskip .2cm
{\it Department of Physics, University of Durham \\
Durham DH1 3LE, England }\\
\vskip   .4cm
and
\vskip .4cm
{\large  W.J. Stirling}
\vskip .2cm
{\it Departments of Physics and Mathematical Sciences, University of Durham \\
Durham DH1 3LE, England }\\
\vskip 1cm
\end{center}
\begin{abstract}
The emission of energetic gluons in $\tt$ production in $\ee$
annihilation can have important experimental consequences, in particular
on top quark mass measurements.
We present compact, analytical expressions for the gluon energy
distribution and its average value at first order in QCD perturbation
theory. Our results are valid for arbitrary masses, collision energies
and production currents. We pay particular attention to top quark
production near threshold, and show that in certain cases the soft
gluon approximation is insufficient to describe the radiation spectrum.

\end{abstract}
\vfill
\end{titlepage}
\newpage
\section{Introduction}
The discovery of the top quark -- one of the basic components of the
Standard Model -- is one of the most important goals for present and
future experiments. Indirect evidence for the existence of top
is very strong, see for example \cite{TOPREVIEW}, and the chances that
it will be detected at the Fermilab $\pp$ collider in the next few years
are very high.
Once top is discovered, the next challenge for experiments is to measure
its parameters -- in particular its mass $M$ -- as precisely as possible.

An unambiguous interpretation of experimental data and the determination of
the top quark parameters relies  on a clear quantitative understanding
of the details of the production process, including the effects of gluon
bremsstrahlung at the production stage \cite{DKT,KOS}.
To quantify these effects, one can study the average amount of energy
lost by the top quark -- the simplest infrared-safe characteristic
of gluon bremsstrahlung. Such a quantity is of potential practical
importance in top mass determinations and, for energies above threshold,
can be calculated quite straightforwardly without undue complications from
non-perturbative phenomena or top-width effects \cite{DKT,KOS,TDK}.
The key point here is the large top width, $\Gamma_t$ \cite{BIGI}. Since
we know from experiment that the top mass is greater than about $120\
\GeV$ \cite{TOPLIMIT}, the decay width is greater than the typical
hadronic scale, $\Gamma_t > \mu \sim 1\ {\rm fm}^{-1}$, and so the top
quark decays before non-perturbative effects like the formation
of $t$-flavoured hadrons become important \cite{BIGI,KUHN}.

The bremsstrahlung of gluons off the top quark can have an
important impact (``radiation damage") on various characterstics
of the $\tt$ final state. Taking $e^+e^-\to\tt$ as an example,
it obviously affects the kinematical constraint  $E_t =
E_{\rm beam}$, over and above the well-known effects of
initial radiation.\footnote{These initial state radiation effects can be
incorporated in a standard way \cite{LK,FK} and will not be addressed
further here.}
It also influences the form of the $\tt $ production vertex and
measurements of the spectra of secondary leptons from $t$-decay
\cite{EE500,FINLAND}.

Since one of the primary goals of a future $\ee$ collider
 would be to obtain an accurate measurement of the top mass $M$, it is worth
 considering the impact  of gluon radiation on such a measurement.
 There are two scenarios which have been considered \cite{IGO}.
 If the top is not too heavy, it can have sizeable kinetic energy
 and a mass measurement can be obtained from the invariant mass
 per hemisphere. In practice, the distributions in the heavy ($M_H$) and
 light ($M_L$) hemisphere masses can be fitted to a Monte Carlo
 simulation with $M$ as a free parameter. However, a gluon radiated
 from the $t\bar t$ system will distort this measurement. In particular
 the average difference in the hemisphere masses squared is proportional
 (at $O(\alpha_s)$) to the average gluon energy,
\beq
\langle M_H^2 - M_L^2 \rangle = \sqrt{s} \langle E_g \rangle \, .
\eeq
The more energy is carried away by the gluon, the more asymmetric
the hemisphere masses. In practice, as we shall see, $\langle E_g \rangle$
can be $O(10\ \GeV)$, which is much larger than  the
estimated  statistical errors on $M$ from such analyses. A Monte Carlo
which does not take full account of gluon radiation  at the $\tt$
production stage could yield misleading results on the precision on $M$.

Another method which is more suited to a heavier top quark is to
reconstruct the mass directly from the three-jet final state \cite{IGO}.
To avoid combinatorial and other backgrounds, the energy of the three-jet
system is typically constrained to be close to the beam energy. But once again,
the emission before the top decay of an energetic gluon can distort
this constraint, since naively
\beq
\langle E_{\rm 3\ jet} \rangle - E_{\rm beam} = - \half \langle E_g \rangle \,
{}.
\eeq
Once again we see the importance of taking the effects of gluon radiation
into account in reconstructing the top mass.

In this paper we present  results for gluon bremsstrahlung in
$\ee\to\tt$ which are of practical interest for physics studies
of future linear $\ee$ colliders. We focus in particular
on the gluon energy spectrum which, as we have seen, can have important
implications for the top  mass measurement.
 It is also straightforward,  at least
in principle, to extend our results to the more complicated case of hadronic
$\tt$ production.

Since centre-of-mass energies $\sqrt{s}$ at which multiple gluon
radiation should be taken into account \cite{DKT,YURI}
 ($\log(\sqrt{s}/M) \gg 1$)
appear unrealistic for the foreseeable future, we need only
consider a first-order perturbative analysis. This is calculationally
quite straightforward, and in fact many parts of the calculation
can be taken over from existing QED analyses \cite{BK1,BK2}
(see also \cite{IOFFE}).

The remainder of the paper is organised as follows.
We first set up the theoretical framework which allows us to
calculate the gluon radiation distribution for arbitrary production currents.
We derive compact, analytical expressions for the gluon energy spectrum
and its average value for vector, axial vector, scalar and pseudoscalar
produciton currents. We then present some numerical calculations which
illustrate the size of the effects for typical experimental parameters.
Finally, we summarize our results and present our conclusions.

\section{Calculation of QCD bremsstrahlung distributions}
In this section we describe the calculation of the cross section for
single primary gluon emission in the process $\ee\to\tt$.
Our calculation is valid for any collision enegy $\sqrt{s} > 2 M$, and
we take both photon and $Z$ exchanges into account.

We begin by defining some variables:
$\sqrt{q^2} \equiv \sqrt{s} = 2E$ is the total centre-of-mass energy and
$M$ is the mass of the top quark.
The momenta of the initial and final state particles are labelled by
$e^-(k_1) + e^+(k_2) \to t(p_1) + \bar{t}(p_2) + g(k)$ and we define
$q^\mu  = p_1^\mu + p_2^\mu + k^\mu$. The energy fractions shared
by the $t$, $\bar t$ and gluon in the $\ee$ centre-of-mass frame
are
\beqn
  z_i &=& {2q\cdot p_i \over q^2} \;, \qquad i=1,2 \nonumber \\
  z   &=& {2q\cdot k   \over q^2} \;, \qquad z  = 2 - z_1 - z_2 \; .
\label{eq1}
\eeqn
The quark velocity in the $\tt$ centre-of-mass frame is
\beq
\beta = \beta(z) \equiv \sqrt{1-\rho_0} = \sqrt{1 - {4\gamma\over 1-z}}
\leq v = \sqrt{1-4\gamma} \; ,
\label{eq2}
\eeq
where
\beq
\gamma = {M^2\over q^2} \leq \frac{1}{4}\; ,\qquad \rho_0 = {4\gamma\over 1-z}
\; .
\label{eq3}
\eeq
The final-state radiation is limited by the maximum kinematically
allowed energy
\beq
0 \leq z \leq z_{\rm max} = v^2 \; .
\label{eq4}
\eeq
The virtualities of the top quark and antiquark before gluon emission,
$\kappa_i$, can be written in terms of the energy fractions:
\beqn
\kappa_1 & = & 2 k\cdot p_1 = q^2 (1-z_2) \; , \nonumber \\
\kappa_2 & = & 2 k\cdot p_2 = q^2 (1-z_1) \; .
\label{eq5}
\eeqn
It is also convenient to introduce the ``angular" variable
\beq
\rho =  {\kappa_1 \over z q^2} = {1-z_2 \over z} \; .
\label{eq6}
\eeq
In the calculations which follow, we neglect the electron mass $m_e$.
Since we are concerned with energy losses caused by gluon emission
at the $\tt$ production stage, we may treat the $t$-quarks as stable
objects. Only very soft gluons with energy $E_g \lapp \Gamma_t E/M$
are affected by the top width, and these make a negligible contribution
to the average energy loss.\footnote{ A detailed discussion of the
potential impact of the top quark instability on gluon emission can be
found in Refs.~\cite{DKT,KOS,TDK,DKOS}.}

\subsection{Lowest-order cross section}
The total non-radiative cross section for $\tt$
production in $\ee$ collisions can be expressed in terms of the vector
and axial contributions (see for example \cite{JERZAK,REVIEWS}):
\beq
\sigma_{\tt} = \sigma^{VV}(s) + \sigma^{AA}(s)
 =  R_{\tt}(s) \sigma_{\rm pt}(s)              \; ,
\label{eq7}
\eeq
where
\beq
\sigma_{\rm pt} (s) = {4\pi\alpha^2(s) \over 3 s }
\eeq
is the point cross section for $\ee \to \mu^+\mu^-$ with the QED
coupling evaluated at the scale $s$, and
\beq
R_{\tt}(s) = R^{VV}(s) + R^{AA}(s) = v\; (\zeta_V  \tau^V(s) + \zeta_A
\tau^A(s) )
\label{eq8}
\eeq
with
\beqn
\label{eq9a}
\zeta_V &=& \half (3-v^2) = 1 + 2\gamma \; , \\
\zeta_A  & =&  v^2 = 1  - 4 \gamma \; .
\label{eq9b}
\eeqn
The electroweak factors $\tau^C$ ($C=A,V$) are defined as
\beqn
\label{eq10}
\tau^V &= & \frac{4}{3} -
 4 v_e v_t \kappa { s(s-M_Z^2 ) \over
(s-M_Z^2)^2 + (s \Gamma_Z/M_Z)^2 }
\nonumber \\
& &  + 3
 (v_e^2 + a_e^2) v_t^2 \kappa^2 {s^2 \over (s-M_Z^2)^2 + (s
\Gamma_Z/M_Z)^2 }   \; ,  \\
\tau^A &= &
3  (v_e^2 + a_e^2) a_t^2  \kappa^2 { s^2 \over
(s-M_Z^2)^2 + (s \Gamma_Z/M_Z)^2 }  \; ,
\label{eq11}
\eeqn
with
\beqn
\kappa = {\sqrt{2} G_F M_Z^2 \over 4 \pi \alpha(s) } \; , \nonumber \\
v_e  = -\half + 2 \sin^2\theta_W\; , & &  a_e = -\half   \; , \nonumber \\
v_t  = \half - \frac{4}{3} \sin^2\theta_W\; , & &  a_t = \half   \; .
\eeqn
The functions $\tau^V(s)$ and $\tau^A(s)$ are shown in Fig.~1.
Except for energies close to the $Z^0$ peak, which are not relevant
for $\tt$ production, the axial current piece is numerically small
compared to the vector current piece.
 At very high energies the ratio of $\tau^A$ to $\tau^V$ tends to a
constant value of approximately 0.26. Furthermore, close to $\tt$ threshold
the axial current contribution is further suppressed by two powers of
$v$, see Eqs.~(\ref{eq9a},\ref{eq9b}).

We should also note that near threshold, the $\tt$ cross section
is strongly modified by Coulomb enhancement effects, see
Refs.~\cite{VSFVAK1,VSFVAK2,PESKIN}.
However the extreme threshold region is not of primary interest in the
present study, since the perturbation radiation is strongly suppressed
there.

\subsection{Dalitz plot distribution for the radiative process}
We consider in this section the calculation of the differential cross
section for the emission of a primary gluon of momentum $k$ in $\tt$
production. After integrating over the relative orientation of the quark
and gluon momenta with respect to the  incoming leptons, the
interference between the vector and axial pieces vanishes and the
double differential cross section in the quark and antiquark energy
fractions (the Dalitz plot  distribution) can be written as
\beq
{d^2 \cW\over dz_1 dz_2 } = {1\over \sigma_{\tt}}\;
{d^2 \sigma_g \over dz_1 dz_2 } =
\frac{R^{VV}}{R_{\tt}} \; \left\{ {d^2 \cW\over dz_1 dz_2 }\right\}_V +
\frac{R^{AA}}{R_{\tt}} \; \left\{ {d^2 \cW\over dz_1 dz_2 }\right\}_A\; ,
\label{eq12}
\eeq
with
\beq
 \left\{ {d^2 \cW\over dz_1 dz_2 }\right\}_C = {1\over \sigma^{CC}}\;
{d^2 \sigma^C_g \over dz_1 dz_2 } \; .
\label{eq13}
\eeq
The subscript $C\ (=V,A)$ denotes the production current, see
Eqs.~(\ref{eq7},\ref{eq8}).

The derivation of the  expressions for the radiative matrix element
squared, summed over colours and spins and integrated over the azimuthal
orientation of the produced particles with respect to the beam
direction, drastically simplifies  if one uses an analogue of the
``invariant-integration technique"
which was employed in Refs.~\cite{BK1,BK2} for calculations of the QED
radiation  accompanying two-charged-particle production in $\ee$
annihilation.

The matrix element $M_C^{(1)}$ describing the radiation off the top
quark produced via the $C$-current exchange can be written as
\beq
M_C^{(1)} = g_s B_C T^a \frac{g_{\mu\nu}}{q^2} \left\{ \bar u(p_1) A^C_\mu
v(p_2) \right\} \cdot  J_\nu^{e,C} \; ,
\label{eq14}
\eeq
with
\beq
 A_\mu^C  = \Gamma_\mu^C (j\cdot e_\lambda)
+ {\hat{e}_\lambda \hat{k} \Gamma_\mu^C \over \kappa_1}
+ {\Gamma_\mu^C \hat{e}_\lambda \hat{k} \over \kappa_2} \; .
\label{eq15}
\eeq
Here $T^a$ is a SU(3) colour matrix and $e_\lambda$ is a polarization
vector for the gluon. The matrices $\Gamma_\mu^C$ ($C=V,A$) are
\beq
\Gamma_\mu^V = \gamma_\mu\; , \quad
\Gamma_\mu^A = \gamma_\mu\gamma_5 \; .
\label{eq16}
\eeq
The four-vector $j_\mu$ is the classical current induced by the
acceleration of the top-quark colour charges,
\beq
j_\mu  = {2 p_{1\mu} \over \kappa_1} - {2 p_{2\mu} \over \kappa_2} \; ,
\label{eq17}
\eeq
and $J^{e,C}_\mu$ describes the $C$-exchange contribution at the
electron vertex,
\beq
J^{e,C}_\mu = \bar v(k_2) \Gamma^C_\mu u(k_1) \; .
\label{eq18}
\eeq
The normalization factors $B_C$ are chosen so
that the non-radiative matrix element $M^{(0)}_C$  is
\beq
M_C^{(0)} =  B_C
\frac{g_{\mu\nu}}{q^2} \left\{ \bar u(p_1)
\Gamma^C_\mu
v(p_2) \right\} \cdot  J_\nu^{e,C} \; .
\label{eq19}
\eeq

Following Refs.~\cite{BK1,BK2}, it is convenient to integrate out the angular
variables which describe the orientation of the final state quarks and
gluons with respect to the incoming leptons. We first introduce the
Lorentz-invariant polarization tensor $\Pi_{\mu\nu}^C$
\beq
\Pi_{\mu\nu}^C   = {1\over 16} \sum_\lambda\; {\rm Tr}\left[
(\hat{p}_1 + M) A_\mu^C (M-\hat{p}_2)\bar{A}^C_\nu \right] \; ,
\label{eq20}
\eeq
with $\bar{A}^C_\nu = \gamma^0 (A_\nu^C)^+\gamma_0$, which describes
the radiative decay of a polarized spin-one heavy
``object" $C$.
Since we wish to integrate over the directions of the quark and
gluon momenta, the resulting tensor structure can depend only on the
four-vector $n_\nu = q_\nu/\sqrt{q^2}$ which defines the reference frame
in which the quark energies are fixed. It is convenient to write the
polarization tensor $\Pi_{\mu\nu}^C$ in the general form
\beq
\Pi_{\mu\nu}^C   =
\frac{1}{3} (g_{\mu\nu} - n_\mu n_\nu ) \Pi_C + n_\mu n_\nu \Pi^q_C \;
,
\label{eq21}
\eeq
where
\beq
\Pi_C   =  \Pi_{\mu\mu} - \Pi^q_C\; ,\quad \Pi^q_C = n_\mu
\Pi^C_{\mu\nu} n_\nu\; .
\label{eq22}
\eeq

Note that because of vector current conservation, the tensor
$\Pi^V_{\mu\nu}$ is transverse, i.e.
\beq
\Pi^q_V = 0\; .
\label{eq23}
\eeq
For the axial current,       the
second (longitudinal) term in Eq.~(\ref{eq21}) vanishes in the limit
$m_e \to 0$. In what follows we will set $m_e = 0$, and so only the
transverse component $\Pi_C$ appears in the final expression for the
double-differential distribution, Eq.~(\ref{eq13}). As a result, we
obtain
\beq
 \left\{ {d^2 \cW\over dz_1 dz_2 }\right\}_C =
\frac{1}{v} \; \frac{C_F\alpha_s}{\pi}\; \Sigma_C \; ,
\label{eq24}
\eeq
where
$$ 
\Sigma_C \equiv \frac{1}4 \, \frac{ \Pi_C}{\Pi_C^{(0)}}
$$
with $\Pi_C^{(0)}$ the non-radiative polarization operator that one
obtains from Eqs.~(\ref{eq21},\ref{eq22}) by substituting the Born
vertices for $A^C$ in Eq.~(\ref{eq20}),
$$
\Pi_{\mu\nu}^{(0)\,C}   = {1\over 16 q^2} \sum_\lambda\; {\rm Tr}\left[
(\hat{p}_1 + M) \Gamma_\mu^C (M-\hat{p}_2)\bar{\Gamma}^C_\nu \right] \; .
$$
With the numerical normalization chosen for Eq.~(\ref{eq20}) one has
\beq
 4\, \Pi_C^{(0)} = \zeta_C \>, \quad
\Sigma_C  =  \frac1{\zeta_C}  \;  {\Pi_C}\>,
\label{eq25}
\eeq
where the Born factors $\zeta_C$ are given in Eqs.~(\ref{eq9a},\ref{eq9b}).
Explicit calculations give\footnote{For completeness, we present in the
Appendix analogous formulae for scalar ($C=S$) and pseudoscalar ($C=P$)
exchanges.}
\beqn
\label{eq26a}
\Pi_V  & = & j^2 \frac{q^2}{4}\, \zeta_V + \frac{1}{2} \left(
\frac{\kappa_2}{\kappa_1} + \frac{\kappa_1}{\kappa_2} \right) \; , \\
\label{eq26b}
\Pi_A  & = & j^2 \frac{q^2}{4} \, \zeta_A + \frac{1}{2} \left[  \left(
\frac{\kappa_2}{\kappa_1} + \frac{\kappa_1}{\kappa_2} \right)(1+2
\gamma)  + 4\gamma \right]
 \; ,
\eeqn
with
\beq
 j^2 \equiv -g^{\mu\nu} j_\mu j_\nu   = \frac{4}{\kappa_1\kappa_2} \left[
(q^2 - \kappa_1 -\kappa_2 - 2M^2) - M^2 \left(
\frac{\kappa_2}{\kappa_1} + \frac{\kappa_1}{\kappa_2} \right) \right] .
\label{eq27}
\eeq
In  terms of the dimensionless energy fractions defined in Eq.~(\ref{eq1}),
we have
\beqn
\frac{q^2}{4} \; j^2  &=& {z_1 + z_2 - 1 - 2\gamma \over (1-z_1)(1-z_2) }
- {\gamma\over (1-z_1)^2 } - {\gamma\over (1-z_2)^2 }
\; ,
\nonumber \\
\frac{\kappa_2}{\kappa_1} + \frac{\kappa_1}{\kappa_2} &=& {z^2 \over
(1-z_1)(1-z_2) }  - 2
\; .
\label{eq28}
\eeqn
Finally, the Dalitz plot distribution is given by substituting
the expressions given in
Eqs.~(\ref{eq8},\ref{eq9a},\ref{eq9b},\ref{eq24} -\ref{eq28}) into
Eqs.~(\ref{eq12},\ref{eq13}).

Before presenting numerical results, we make some general comments
concerning the global structure of the above results for the $\Pi_C$.
\begin{itemize}
\item [{(i)}]
The first (\lq\lq classical") term in Eqs.~(\ref{eq26a},\ref{eq26b})
corresponds to long-distance radiation with polarization vector
$\vec{e}_\parallel$ in the plane defined by $\vec{p}_1, \vec{p}_2$.
It leads to a universal contribution to the radiative cross section.
In contrast, the short-distance effects are represented by the second
term, which induces a current-dependent, non-universal contribution.
This describes equal production of the gluon  states $\vec{e}_\parallel$
and $\vec{e}_\perp$ (polarization transverse to the $\vec{p}_1, \vec{p}_2$
plane).
\item [{(ii)}] Next, consider the case of soft gluon emission, when the
radiative cross  section can be expanded in powers of the gluon energy
fraction $z \ll 1$. In accordance with the factorization properties of
soft radiation \cite{XXXX}, the first term in Eqs.~(\ref{eq26a},\ref{eq26b})
is proportional to the product of the Born term $\zeta_C$ and the usual
accompanying radiation factor $j^2$, with the former evaluated
at the total centre-of-mass energy $q^2$ of the process.
It is interesting that at small $z$ the second (non-universal) term
in Eqs.~(\ref{eq26a},\ref{eq26b}) appears to be $O(z^2)$ compared to the
first term, not $O(z)$ as one might naively expect. The only correction
linear in $z$ is incorporated into the classical radiation factor
$j^2$. The physical origin of this can be understood \cite{DK} by
recalling the celebrated Low soft bremsstrahlung theorem \cite{LOW},
extended to the case of charged fermions by Burnett and Kroll
\cite{BURNETT}. The application of these classical results is especially
transparent for the case of the integrated annihilation cross sections,
where the non-radiative process depends only on one energy variable (the
centre-of-mass energy of the charged particles).\footnote{The
Low-Kroll-Burnett approach leads to simple results also for
initial-state radiation in annihilation processes. For example, in
$\ee$ annihilation, the universal part of the integrated photon-emission
cross section is given by the product of the Born term
taken at the point $(q-k)^2$ and the radiation factor describing the
emission off the initial electrons \cite{BK1,BK2,DK}. This construction
remains valid for the more general case of arbitrarily polarized,
massive initial-state particles (see also \cite{BK3}).}
\item [{(iii)}]
In the extreme ultra-relativistic limit, $\gamma \ll 1$,
the $\Pi_V$ and $\Pi_A$ distributions become identical.
\item [{(iv)}]
The results (\ref{eq24}-\ref{eq26b}) are not valid in the extreme
threshold limit, even within the context of perturbation theory.
In this region, the first-order radiation amplitude
is strongly modified by the QCD Coulomb-like interaction between the
$t$ and $\bar t$ \cite{VSFVAK2}. This leads to a distortion of the radiation
pattern and to the additional suppression of radiation
\cite{VSFVAK2,FADYAK}.
\end{itemize}

\subsection{Gluon energy spectrum}
To calculate the inclusive gluon energy distribution it is convenient to
rewrite  the quantities $\Sigma_C$ in terms of the variables $z$
and $\rho$ (see Eqs.~(\ref{eq6},\ref{eq25}-\ref{eq26b})). This gives
\beqn
\Sigma_V & = & \Sigma_{\rm soft}  + \half z^2 \frac{1}{\zeta_V}
\left[ {1\over \rho} -1 \right] \; + \; (1\leftrightarrow 2) \; ,
\label{eq29a}
\\
\Sigma_A & = & \Sigma_{\rm soft}  + \half z^2 \frac{1}{\zeta_A}
\left[ {1\over \rho} -1  + {2\gamma\over \rho} \right] \; + \;
(1\leftrightarrow 2) \; .
\label{eq29b}
\eeqn
Here we have introduced $\Sigma_{\rm soft}$ as the classical current
contribution which accounts for the $z^0$ and $z^1$ terms,
\beq
\Sigma_{\rm soft} = {1-z-2\gamma \over \rho} - {\gamma\over \rho^2} \; ,
\label{eq30}
\eeq
and
\beq
\half (1-\beta(z))   \leq \rho \leq \half (1+\beta(z))
\; .
\label{eq31}
\eeq
Explicit integration then gives
\beq
z \left\{ {d \cW\over dz }\right\}_C = {z\over \sigma^{CC}}\;
{d^2 \sigma^C_g \over dz}  =
{\beta(z)\over  v} {C_F\alpha_s\over \pi} \left[
\xi(z) + \zeta_C^{-1} z^2 \xi_H(z;C) \right]     \; ,
\label{eq32}
\eeq
with the Born factors $\zeta_C$ given in Eqs.~(\ref{eq9a},\ref{eq9b}).
The universal \lq\lq soft" part of the gluon density  reads
\beq
\xi(z) = (1+v^2 -2 z)\cL(z) - 2(1-z) \; ,
\label{eq33}
\eeq
where
\beq
\cL(z) = \frac{1}{\beta(z)} \ln{1+\beta(z)\over 1-\beta(z)} \; .
\nonumber
\eeq
The \lq\lq hard" contributions which depend on the the production
channel are
\beqn
\xi_H(z;V)  &=& \cL(z) - 1
\; , \nonumber \\
\xi_H(z;A)  &=& \cL(z)\left[{3-v^2 \over 2} \right]  - 1
\; .
\label{eq34}
\eeqn
Note that the result for $ \left\{ {d \cW\over dz }\right\}_V$
coincides with that obtained in Ref.~\cite{BK1}.

\subsection{Energy integrals}
We turn now to the quantity which is the main concern of the present
study -- the average energy lost by the $t$-quark by gluon emission
at the production stage.

The average gluon energy fraction is
\beq
\av{z}_C = \int_0^{v^2} dz \; z \; \left\{ {d \cW \over d z} \right\}_C
\label{eq35}
\eeq
and the mean energy fraction lost by the $t$-quark (antiquark)
is then
\beq
\av{z_1} = \av{z_2} = 1 - \half \av{z}_C \; .
\eeq
It is convenient to write $\av{z}_C$ in the form
\beq
\av{z}_C  = C_F\; {\alpha_s\over\pi}\; \left\{ \cZ + \zeta_C^{-1}
\cZ_C\right\} \; ,
\label{eq37}
\eeq
where the universal $C$-independent piece is
\beq
\cZ = {2\over v}\; \int_0^{v^2} dz\; \beta(z)\; [(1-z -2\gamma)\cL(z)
-(1-z) ] \; .
\label{eq38}
\eeq
The \lq\lq hard" term $ \cZ_C$ in Eq.~(\ref{eq37}) is given by
\beqn
\cZ_V &=& I_{\cL} - I_1 \; , \nonumber \\
\cZ_A &=& I_{\cL}\; \left[ {3-v^2 \over 2} \right]  - I_1 \; ,
\label{eq39}
\eeqn
with
\beqn
I_{\cL}  &=& {1\over v} \int_0^{v^2}dz \; z^2 \;  \beta(z)\; \cL(z)
=   {1\over v}  \int_0^{v^2} dz\; z^2\; \ln{1+\beta(z)\over
1-\beta(z)}\; ,
\nonumber \\
I_{1}  &=& {1\over v} \int_0^{v^2}dz \;z^2 \; \beta(z)\; .
\label{eq40}
\eeqn
Performing the $z$ integration in Eqs.~(\ref{eq38},\ref{eq40}), we obtain
\beqn
 \cZ &=& {1\over 8}\left\{   (3+2v^2+3v^4)\cL_v -6(1+v^2) \right\}\; , \\
I_{\cL} &=& {1\over 16}\left\{ \left({5\over 3} + v^2 + v^4 +
{5v^6\over 3} \right)\cL_v
-2\left({5\over 3} + {14v^2\over 9} + {5v^4\over 3} \right) \right\}\; , \\
 I_{1} &=&  {1\over 16}\left\{ \left(-5+ 3v^2+ v^4+ v^6\right)\cL_v
+2\left(5 - {4v^2\over 3}  -v^4 \right)  \right\} \; ,
\eeqn
with
\beq
 \cL_v \equiv \frac1v \ln\frac{1+v}{1-v}\; .
\label{eq42}
\eeq

The average gluon energy fractions for vector and axial
production currents are  obtained by combining the results
of Eqs.~(\ref{eq35}-\ref{eq42}). Figure~2 shows the
resulting  $\av{z}_V$ and
$\av{z}_A$ as a function of the top quark velocity $v$, for
$\alpha_s = 0.1$.
Note that in the relativistic limit, $v\to 1$, the average gluon energies
diverge logarithmically,
\beq
\av{z}_V \approx \av{z}_A \approx
 C_F\; {\alpha_s\over\pi}\; \left[ \frac{4}{3} \;
\ln\frac{2}{1-v}  \; - \; \frac{22}{9} \right] \; ,
\label{vto1}
\eeq
reflecting the emergence of the collinear singularity  in the
massless quark limit. Of course when $\alpha_s \ln(1/(1-v)) = O(1)$
multiple gluon emission becomes important, and the leading logarithms
must be resummed to all orders (see Ref.~\cite{DKT}).
The behaviour of the average energy in the threshold region, $v\to 0$,
is discussed in the following section.

Figure~3 shows that average gluon energy $\langle E_g \rangle =
\langle z \rangle \sqrt{s} / 2$ as a function of the top mass $M$
in $\ee$ annihilation at centre-of-mass energy $\sqrt{s}$. The correct
proportions of vector and axial current contributions (Fig.~1) are included,
and the strong coupling is taken to be $\alpha_s = 0.1$. For top masses
in  the experimentally favoured range $120 - 180$ GeV,  $\langle E_g \rangle$
varies from 14 GeV to 4 GeV.

\subsection{Non-relativistic case}

For $v^2 \ll 1$, we can take the small-$v$ limit of the results
of the previous section to obtain
\beqn
\label{eq43a}
\cZ & \simeq & \frac{16}{15}\; v^4\; \left[ 1 + O(v^2) \right] \; , \\
\label{eq43b}
I_{\cL} & \simeq & \frac{2v^2}{7} \; \cZ \; , \\
\label{eq43c}
I_1 & \simeq & \frac{v^2}{7} \; \cZ \; .
\eeqn
The high powers of the threshold factor $v$ originate as follows: in the
case of the ``classical" universal term $\cZ$, one power of $v^2$ comes
from  the size of the integration region, $z_{\rm max}=v^2$ (Eq.~(\ref{eq4})),
and another power comes from the dipole suppression of the accompanying
radiation, see also Refs.\cite{BK2,VSFVAK2}. Because of the
explicit $z^2$ factor appearing in the ``hard radiation" spectrum,
Eq.~(\ref{eq32}), the non-universal contribution $\cZ_C \sim z_{\rm max}^3
  = v^6$, and
thus acquires an additional $v^2$ suppression relative to the universal
$\cZ$ piece. Note, however, that the impact of the ``hard radiation" in
the threshold region strongly depends on the production channel. In
particular, the Born factor $\zeta_C$ in the denominator in
Eq.~(\ref{eq37}) can compensate the suppression of the non-classical
effects in  $\cZ_C$, see Eq.~(\ref{eq9a},\ref{eq9b}). This is an interesting
example
of how the universal nature of soft radiation can sometimes be obscured by
process-dependent short-distance effects.

In the vector channel, where the ($S$-wave) Born amplitude is not
suppressed at threshold, the soft piece dominates and we find
(see also \cite{BK2,VSFVAK2})
\beq
\av{z}_V  \sim C_F\; {\alpha_s\over\pi}\; \frac{16}{15} \;
v^4\; \left[ 1 + O(v^2) \right] \; .
\label{eq44}
\eeq
However, in the case of the  axial-vector ($P$-wave) production channel,
the non-radiative amplitude for the $C\to\qq$ decay is additionally
suppressed at threshold.
As discussed above, ``non-classical" radiation effects at $O(\alpha_s)$
are then equally important, and induce
a significant
contribution to $\av{z}_A$.
 From Eqs.~(\ref{eq37},\ref{eq39},\ref{eq43a}-\ref{eq43c}) we find
\beq
\av{z}_A \sim C_F\; {\alpha_s\over\pi}\; \frac{16}{15} \;
v^4\; \left( 1 + \frac{2}{7}\right)\; \left[ 1 + O(v^2) \right] \; ,
\label{eq45}
\eeq
with the second term in the $()$ brackets coming from the
``hard radiation". The limiting $v\to 0$ behaviours corresponding to the
leading $v^4$ terms in $\av{z}_V$ and $\av{z}_A$ are shown as dashed
lines in Fig.~2.

As we have already explained, the practical consequences of the
differences between the average gluon energy in the vector and
axial-vector channels for $\ee\to\tt$ are not large, since the
axial contribution is so small in the threshold region, see
Eqs.~(\ref{eq8},\ref{eq9a},\ref{eq9b},\ref{eq12}).
It is, however, worth mentioning
that the non-universality of the soft radiation in the threshold region
is not restricted to the axial channel. The same situation arises
in the case of  $\tt$ production in the scalar (e.g. Higgs exchange)
channel, $S\to\qq$.  In the Appendix we present the corresponding
results for $\av{z}_S$. We find that in the limit $v\to 0$ we have
$\av{z}_S \simeq \av{z}_A$, so that once again  the soft
radiation contribution is accompanied by a hard radiation term
of the same order.

Another example which could be of practical interest concerns gluon
radiation accompanying two scalar quark (e.g. stop) production:
\beq
\ee\to \tilde{q}\tilde{\bar q} \; .
\eeq
This is a $P$-wave process, dominated by  photon
exchange.\footnote{Detailed QED results for radiation accompanying two charged
scalar particle production can be found in Ref.~\cite{BK2}.}  Near
threshold, the average energy loss is again given by Eq.~(\ref{eq45}),
illustrating the non-universality of the  soft radiation
description   for this process also.

All the  results derived above have assumed that the top
is a {\it stable} particle. However in practice and
particularly near threshold, one must
be careful to take account of the effects of the top decay width
and of non-perturbative fragmentation. Although the former
can affect the distribution of soft gluons (i.e. gluons
with energy $E_g \lapp \Gamma_t$) \cite{DKT,KOS}, it is not important for
the average top energy loss.
We can quantify the possible impact of non-perturbative dynamics
by using the string-model results of Ref.~\cite{SJOZER} for the
average amount of energy lost by the top quark before it decays. Near
the $\tt$ threshold this leads to
\beq
\av{z}_{\rm non\; PT} =  {2 \kappa \hbar \over \Gamma_t M} \; v\; ,
\label{eq47}
\eeq
with the so-called string tension factor $\kappa \simeq 1\ \GeV/{\rm
fm}$. Since $\Gamma_t \sim M^3$ \cite{BIGI}, these non-perturbative
effects decrease as $M^{-4}$. Taking as a specific example the canonical
values $M = 150\ \GeV$ and  $\Gamma_t = 0.8\ \GeV$, we obtain
\beq
\av{\Delta z}_{\rm non\; PT} \simeq 3.3\times 10^{-3} \; v \; .
\label{eq48}
\eeq
Comparing Eqs.~(\ref{eq44}) and (\ref{eq48}), one may  conclude
that the perturbative prediction is only valid for $v^2 > 0.17$.

\section{Conclusions}

It is important that gluon emission in top quark production processes
is under control. Otherwise, it may provide a serious source of
uncertainty in the determination of the top quark parameters from
experimental data.
In this paper we have studied gluon bremsstrahlung  off the top quarks
in $\ee\to\tt$. We have calculated
the exact first-order perturbative expressions for the energy fraction
distributions of the $t$-quarks. These are valid in a wide energy range
above the $\tt$ threshold.

To examine the impact of gluon emission on top quark distributions we
have calculated the average energy loss at the production stage. This
infra-red safe quantity characterizes, for example, the difference
between the beam energy and the actual energy shared by the produced
$t$-quarks.

\bigskip
\medskip
\noindent{\Large\bf Acknowledgements}
\bigskip

\noindent We are grateful to Victor Fadin, David Miller, Lynne Orr
and Torbj\"orn Sj\"ostrand for useful discussions.
This work was supported in part by
the United Kingdom Science and Engineering Research  Council.

\newpage
\setcounter{section}{1}
\setcounter{equation}{0}
\renewcommand{\thesection}{\Alph{section}}
\renewcommand{\theequation}{\Alph{section}\arabic{equation}}
\section*{Appendix}

In this Appendix we present the
expressions corresponding to single primary gluon emission in the
process $\ee\to\qq$, mediated by the exchange of
scalar $S$  or pseudoscalar $P$ particles (for example, the Higgs
bosons of the Standard Model or its supersymmetric extensions).
Now the invariant polarization operator bears no Lorentz indices,
and we define (cf. Eq.~(\ref{eq20}))
\beq
\Pi_C   = {1\over 8} \sum_\lambda\; {\rm Tr}\left[ (\hat{p}_1
+ M) A^C (M-\hat{p}_2)\bar{A}^C \right] \; ,
\label{alaeq20}
\eeq
where the normalization has been fixed to preserve the relation (\ref{eq25}).
Here the scalar functions $A^C$ are analogous to $A_\mu^C$ of
Eq.~(\ref{eq15}) but with the vertex operators
\beq
\Gamma^S = 1\; , \quad
\Gamma^P = \gamma_5 \; ,
\label{alaeq16}
\eeq
substituted for $\Gamma_\mu^C$.

Defining the double-differential distribution analogously to
Eqs.~(\ref{eq24}), (\ref{eq25}) with
\beq
\zeta_S = \zeta_A = v^2\; , \quad \zeta_P = 1 \; ,
\eeq

we find
\beqn
\Pi_S &=& j^2\; {q^2\over 4}\; v^2 \; + \; {1\over 2}\left(
{\kappa_2\over \kappa_1} + {\kappa_1\over \kappa_2} +  2 \right) \; , \\
\Pi_P &=& j^2\; {q^2\over 4}\;  + \; {1\over 2}\left(
{\kappa_2\over \kappa_1} + {\kappa_1\over \kappa_2} +  2 \right) \; .
\eeqn
The gluon energy spectrum is then given by Eq.~(\ref{eq32})
with
\beq
\xi_H(z,S) = \xi_H(z;P) = \cL(z)\; ,
\eeq
and the mean gluon energy fraction by Eq.~(\ref{eq37}) with
\beq
\cZ_S = \cZ_P  = I_{\cL} \; .
\eeq
Near the threshold for the pseudoscalar channel, the  soft term
dominates and
\beq
\av{z}_P \simeq \av{z}_V \; .
\eeq
In contrast, the transition $O^+ \to \qq$ is $P$-wave, and the Born
term $\zeta_S = v^2$ compensates the additional suppression of $\cZ_S$.
The threshold result for     $\av{z}_S $ therefore
coincides with that for $\av{z}_A$,  Eqs.~(\ref{eq43a}-\ref{eq43c}).

\newpage
\section*{Figure Captions}
\begin{itemize}
\item [{[1]}]
The functions $\tau^V(s)$ and $\tau^A(s)$ defined in
Eqs.~(\ref{eq10},\ref{eq11}).
\item [{[2]}]
The average gluon energy in $e^+e^-\to \tt g$ as a function of the
top quark velocity $v$, for vector and axial vector current production.
Also shown (dashed lines) are the limiting $v\to 0$ behaviours given
in Eqs.~(\ref{eq44},\ref{eq45}), and the limiting $v\to 1$ behaviour given in
Eq.~(\ref{vto1}).
\item [{[3]}]
The average gluon energy in $\tt $ production in $\ee$  annihilation
at $\sqrt{s} = 500\ \GeV$, as a function of the top quark mass $M$.
The strong coupling is fxed at $\alpha_s = 0.1$.

\end{itemize}
\newpage

{\Large\bf Fig.~1}

\vskip  1truecm
{\Large\bf Fig.~2}

\vskip  1truecm
{\Large\bf Fig.~3}

\vskip  1truecm

\end{document}